# Effective Sorting of Fractional Optical Vortex Modes

Zhengyang Mao,[1,*] Haigang Liu,[1,*,†] and Xianfeng Chen[1,2,3,‡]

[1]*State Key Laboratory of Advanced Optical Communication Systems and Networks, School of Physics and Astronomy, Shanghai Jiao Tong University, Shanghai 200240, China*
[2]*Shanghai Research Center for Quantum Sciences, Shanghai 201315, China*

Mode sorter is the crucial component of the communication systems based on orbital angular momentum (OAM). However, schemes proposed so far can only effectively sort integer OAM (IOAM) modes. Here, we demonstrate the effective sorting of fractional OAM (FOAM) modes by utilizing the coordinate transformation method, which can convert FOAM modes to IOAM modes. The transformed IOAM modes are subsequently sorted by using a mode conversion method called topological charge matching. The validation of our scheme is verified by implementing two sorting processes and corresponding mode purity analyses, both theoretically and experimentally. This new sorting method exhibits a huge potential of implementing a highly confidential and high-capacity FOAM-based communication system, which may inspire further applications in both classical and quantum regimes.

Angular momentum is one of the most vital and fundamental properties of physical particles [1-3]. Since Allen discovered that vortex beams with a helical wavefront phase distribution $exp\,(i\ell\theta)$ carry orbital angular momentum (OAM) in 1992 [2], where $\theta$ is the azimuthal coordinate and $\ell$ is the topological charge, researchers have conducted thorough studies on vortex beams for years, demonstrating its applications in optical manipulation [4-6], micromachining [7] and super-resolution imaging [8]. The unlimited range of $\ell$ implies that OAM beam possesses an infinite number of eigenstates that can be exploited to infinitely elevate the capacity of communication system [9-11].

Due to its significant importance, varieties of techniques have been proposed for the generation of vortex beams, ranging from diffractive optical elements [12-14], spatial light modulators [15] to metasurfaces [16-19]. Additionally, in the OAM-based communication systems such as OAM shift keying (OAM-SK) [20] and OAM division multiplexing (OAM-DM) [9], a mode sorter capable of effectively sorting different OAM modes is particularly crucial. Therefore, there has been enormous enthusiasm on sorting of OAM. Some accomplished schemes contain interference, diffraction and mode conversion methods [21,22]. Sorting techniques based on Mach-Zehnder interferometer can separate even and odd OAM modes into two ports [23,24]. And one optical coordinate transformation method that transforms azimuthal phase to horizontal phase provides a simple and efficient approach to sort vortex beams [25,26]. In addition, metasurfaces designed for the sorting of optical OAM modes have also been developed [27,28]. All those methods have propelled the development of sorting techniques for optical vortex modes.

However, in the aforementioned methods, the effective sorting can only be applied to input modes with integer OAM (IOAM). Actually, $\ell$ can also be a non-integer, corresponding to the fractional OAM (FOAM) beam [29-33]. In 2004, Berry provided the first theoretical analysis of the FOAM mode, proving that it could be regarded as a superposition of IOAM eigenmodes [29]. Based on its novel properties markedly different from IOAM mode [34,35], FOAM mode has further expanded the scope of OAM applications, achieving unique functionalities in multiple domains, including the precise control over cell orientation in optical micro-manipulation [36,37] and the anisotropic edge enhancement in optical imaging [38-40]. Another promising application is the optical communication system based on FOAM modes, which holds the promise in further expanding the communication capacity [41,42]. But the lack of effective FOAM mode sorter hinders its further development. One method utilizing the machine learning algorithms enables the high-resolution identification rather than the effective sorting of FOAM modes [43]. Therefore, it is of fundamental importance to develop effective sorting technique applicable to FOAM input modes.

In this paper, we propose a method for the effective sorting of fractional vortex modes. Following the coordinate transformation method, we first convert the FOAM non-eigenmodes into IOAM eigenmodes. Subsequently, the transformed IOAM beam is effectively sorted using a mode conversion method called topological charge matching, which is commonly used in OAM-based optical communication [9]. To evaluate its performance, we conduct sorting experiments on different groups of FOAM modes and perform the purity analysis on the sorted FOAM modes. The experimental > 86% purity verifies the success of our new approach in effective sorting of fractional vortex modes.

The coordinate transformation theory in ray optics is the foundation of our technique, which enables specific mapping operations on the spatial position of light. Consider the paraxial propagation of a parallel light between two planes, where there is a distance d between the input plane $(x, y)$ and the output plane $(u, v)$. The Generalized Snell's law tells that there exists a coordinate transformation between points on two planes, which is described by the formulas [44]

$$Q_x = k\frac{u-x}{d}, Q_y = k\frac{v-y}{d}, \qquad (1)$$

where $k$ is the wave vector in free space and $Q(x, y)$ is the transformation phase loaded on the input plane. By loading different $Q(x, y)$, various types of coordinate transformation can be achieved. Spiral transformation is a special transformation from one spiral to another. The input and output planes are redefined in terms of spiral-polar coordinate described by the radial coordinate $r$ and the spiral azimuthal coordinate $\theta$ varies in $[0, +\infty)$. The entire plane can be partitioned by a set of spirals with no $2\pi$ restriction on angular coordinates, which allows for finer divisions of the plane.

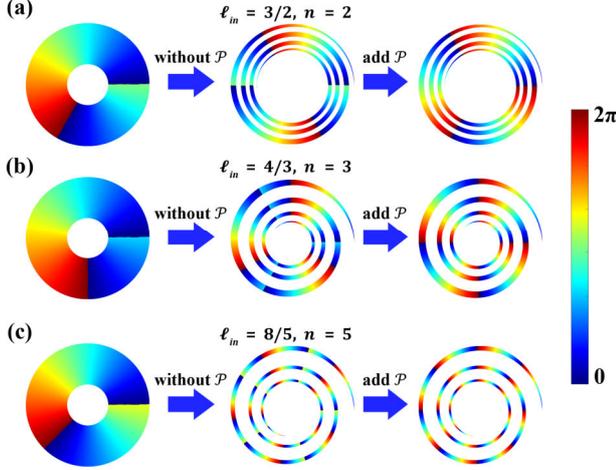

FIG. 1. Theoretical prediction of the phase distribution of the input and output OAM beams with (a) $\ell_{in} = 3/2$ and $n = 2$, (b) $\ell_{in} = 4/3$ and $n = 3$, (c) $\ell_{in} = 8/5$ and $n = 5$. The second and the third column illustrates the phase distribution of the output beam after the transformation without and with $\mathcal{P}(\rho, \varphi)$, respectively.

The specific transformation laws for spiral transformation are described by [44]

$$\rho = cr^{-1/n}, \varphi = \theta/n, \quad (2)$$

where $(r, \theta)$ and $(\rho, \varphi)$ are the coordinates of the input and output planes, $c$ is an arbitrary constant and $n$ is the transformation factor. Assuming that $(r, \theta)$ lies on a spiral defined by $r = ae^{b\theta}$ which is used in our method, according to Eq. (2), $(\rho, \varphi)$ will also lie on a new spiral $\rho = ca^{-1/n}e^{-b\varphi}$. Based on Eq. (1) and (2), the required transformation phase $Q_1(r, \theta)$ on input plane is calculated as

$$Q_1(r, \theta) = \frac{k}{d}\left[\frac{cr^{1-1/n}}{1 - 1/n}\cos(\theta - \theta/n) - \frac{r^2}{2}\right]. \quad (3)$$

Another phase mask $Q_2(\rho, \varphi)$ (correction phase) on the output plane is required to compensate for both $Q_1(r, \theta)$ and the propagation phase generated during the propagation in free space, which is written as

$$Q_2(\rho, \varphi) = -Q_1 - k\sqrt{r^2 + \rho^2 - 2r\rho\cos(\varphi - \theta) + d^2}. \quad (4)$$

For FOAM input modes, an additional correction phase $\mathcal{P}$ is superimposed onto $Q_2$, which is expressed as

$$\mathcal{P}(\rho, \varphi) = 2\pi t\left[\frac{n\varphi}{2\pi}\right], \quad (5)$$

where $t = mod(\ell_{in}, 1)$. Such a requirement stems from the unique characteristic of the FOAM modes that its phase distribution in the azimuthal direction is discontinuous. The spiral transformation compresses or stretches the input optical field by the factor of $n$ along the azimuthal direction without directly changing the phase. The discontinuities in azimuthal phase persist after spiral transformation, leading to the azimuthal phase jump on the output beam, for which the correction phase $\mathcal{P}$ is introduced to compensate.

If the incident beam is a Gaussian vortex beam with the waist radius $w_0$ and a topological charge $\ell_{in}$, the amplitude of the field before and after the transformation is given by

$$E_1(r, \theta) = \frac{1}{w_0}\left(\frac{r\sqrt{2}}{w_0}\right)^{|\ell|} exp\left(-\frac{r^2}{w_0^2}\right)exp(i\ell_{in}\theta), \quad (6)$$

$$E_2(\rho, \varphi) = -iexp(ikd)|n|\frac{r}{\rho}E_1(r, \theta). \quad (7)$$

The exponential term in Eq. (7) is $exp(i\ell_{in}\theta) = exp(in\ell_{in}\varphi)$, which indicates that the transformed beam is a vortex beam carrying OAM with $\ell_{out} = n\ell_{in}$. Figure 1 present the theoretical prediction of the phase distribution of the optical field before and after our spiral transformation with different FOAM input modes $\ell_{in}$ and transformation factors $n$. The second and the third column of Figs. 1(a)-1(c) illustrates the phase distribution of the output beam after the transformation without and with $\mathcal{P}$, respectively. The phase distributions of the transformed beam without $\mathcal{P}$ possess azimuthal jumps in the horizonal direction, which are entirely different from the intended integer vortex modes. And the results with $\mathcal{P}$ demonstrate the successful conversion from FOAM modes to target IOAM modes. It is worth mentioning that, $\mathcal{P}$ remains identical for the same $t$, indicating the generality of our method to be applied to a set of FOAM modes with the same $t$, such as all half-integer FOAM modes.

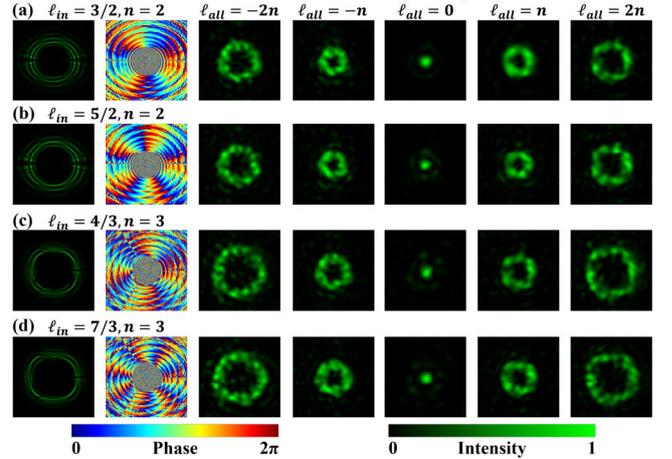

FIG. 2. Numerical results for FOAM sorting. The first two columns are the intensity and phase distribution of the output beam. The last five columns are the results of topological matching for $\ell_{all} = -2n, -n, 0, n, 2n$. (a) $\ell_{in} = 3/2$ and $n = 2$. (b) $\ell_{in} = 5/2$ and $n = 2$. (c) $\ell_{in} = 4/3$ and $n = 3$. (d) $\ell_{in} = 7/3$ and $n = 3$.

The output vortex beam obtained after spiral transformation will be sorted by topological charge matching. The concrete operation is applying a vortex phase map with a topological charge of $\ell_m$ to the output beam, after which the amplitude becomes

$$E_3(\rho, \varphi) = E_2(\rho, \varphi) \times exp(i\ell_m\varphi) \\ \propto exp(i(\ell_{out} + \ell_m)\varphi). \quad (8)$$

Obviously, when the topological charges match each other $(\ell_{out} + \ell_m = \ell_{all} = 0)$, the exponential term representing OAM will be completely eliminated, which means that after focusing through a lens, the intensity distribution on the focal plane will revert back from a halo to a Gaussian-like spot in the center of the field. In cases of $\ell_{all} \neq 0$, the halo still exists. Therefore, we can sort the target vortex mode of $\ell_{in}$ using a spatial filter to extract the central spot after adding the vortex phase map of $\ell_m = -\ell_{out} = -n\ell_{in}$. Since the FOAM modes are not the eigenstates, the direct use of topological charge matching method will cause strong crosstalk [45]. After our coordinate transformation, the FOAM modes are transformed into IOAM eigenmodes, thereby enabling effective sorting such FOAM states with low crosstalk.

The numerical simulations of the effective sorting of FOAM modes are performed by rigorously computing the Fresnel integral on the incident light field, which are shown in Fig. (2). The parameters involved in the simulations are $d = 1\ cm$, $k = 2\pi/\lambda$, $\lambda = 532\ nm$, $r_0 = 240\ \mu m$, $w_0 = 100\ \mu m$, $b = 0.05$ and $c = r_0^{1+1/n}$. Figures 2(a)-2(d) illustrate the numerical results of the intensity distribution, phase distribution and topological charge matching of the transformed beam on the output plane. From top to bottom, the input topological charges are $\ell_{in} = 3/2, 5/2, 4/3, 7/3$ and the transformation scaling factors are $n = 2, 2, 3, 3$, respectively. The corresponding target IOAM modes are $\ell_{out} = 3, 5, 4, 7$. The first two columns are the intensity distribution and the phase distribution of the transformed beam while the last five columns are the results of topological charge matching. To better demonstrate the functionality of topological charge matching in sorting vortex modes, we simulated the results under conditions of both matching ($\ell_{all} = 0$) and mismatching ($\ell_{all} = \pm n, \pm 2n$). Therefore, $\ell_{all} = \pm n, \pm 2n$ is used to simulate condition of mismatching. There is no need to use other $\ell_m$ since there are no matching IOAM modes. The condition of $\ell_{all} = 0$ means the topological charge of the added vortex phase map is the opposite of the target IOAM mode ($\ell_m = -n\ell_{in}$). Therefore, we can determine the success of the sorting based on the presence of a single Gaussian spot appears in the center of the light field under the condition of $\ell_{all} = 0$.

The corresponding numerical results align well to our theoretical predictions. As shown in Fig. (2), the intensity distributions exhibit a spiral vortex pattern. There are several intensity breakpoints on the spiral, which originate from the radial opening ring pattern of the FOAM modes. One can clearly see that the phase distribution exhibits a good continuity with no azimuthal phase jumps, demonstrating continuous cyclic variations from 0 to $2\pi$ that is repeated $n\ell_{in}$ times, which is exactly consistent with the target IOAM mode. The successful transformation from FOAM to IOAM mode is further validated by the results of topological charge matching. It is evident that for all the FOAM input modes, the intensity distribution of the output field is a single Gaussian-like spot in the center of the field when $\ell_{all} = 0$. And the result for $\ell_{all} = \pm n, \pm 2n$ is a light ring significantly larger than the spot. There is almost no overlapping region between the spot and the light ring, indicating that the crosstalk between different FOAM modes is very low when using spatial filter for sorting. Such simulation results confirm the validity of our scheme to effectively sort the FOAM modes.

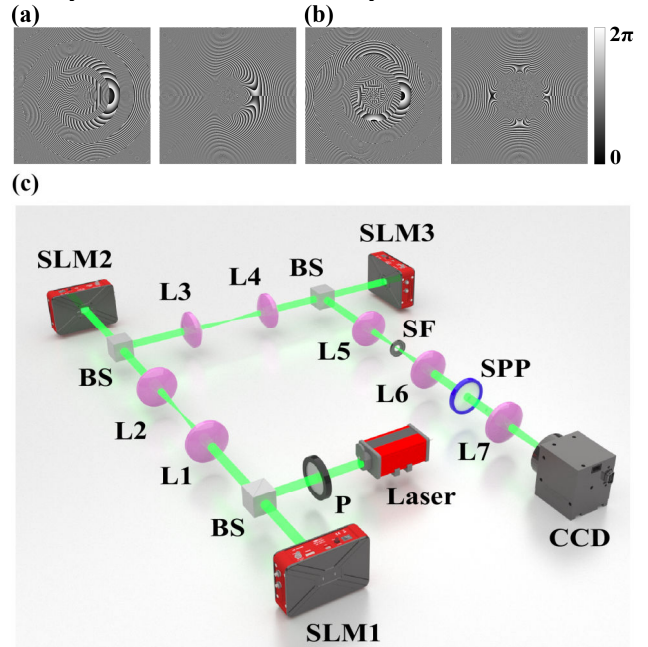

FIG. 3. (a)-(b) Phase maps with different parameters that are used in our experiment. The transformation phase and the correction phase are shown from left to right, respectively. (a) $t = 1/2$ and $n = 2$. (b) $t = 1/3$ and $n = 3$. (c) Schematic of the experimental setup. BS, beam splitter (see the text for more details).

Furthermore, we carry out the experiment to verify the theoretical results above. The phase modulation of the FOAM mode is achieved by using a spatial light modulator (SLM). The phase maps loaded onto the SLM are calculated with the formulas mentioned above. Specifically, the transformation phase map is obtained by $Q_1$ and the correction phase map is calculated using $Q_2 + \mathcal{P}$. The parameters used in the calculations remain consistent with those employed in simulations. The results are converted into grayscale images suitable for SLM. Figures 3(a)-3(b) present the phase maps that are applied to experiment and the transformation phase and the correction phase are presented from left to right.

The experimental setup is illustrated in Fig. 3(c). The light source is a continuous wave (CW) laser with the wavelength of 532 nm. The polarization of the laser beam is changed to horizontal direction by using the polarizer (P). Subsequently, the laser beam is reflected by SLM1, which is loaded with a vortex phase map to produce input FOAM states. The SLM we use is a reflective phase-only liquid crystal modulator (UPOLabs, HDSLM80R), which has a rate of 60 Hz and a resolution of $1920 \times 1080$ pixels. The input light is illuminated onto SLM2 loaded with a transformation phase map via a $4f$ system composed of lenses L1 and L2. The focal lengths of the lenses used in the experiment are all $f = 100\ mm$. The modulated beam is directed onto SLM3 loaded with the correction phase through another $4f$ system consisting of L3 and L4, where SLM3 is located on the back focal plane of this $4f$ system. There is a distance of $f + d$ between SLM2 and L3, which ensures the free propagation of light between the input and output plane at a distance of $d$, as required by the coordinate transformation principle. A low-pass spectral filtering process is implemented via a third $4f$ system with a spatial filter (SF) in the intermediate Fourier plane to isolate the desired IOAM beam. A spiral phase plate (SPP) with topological charge $\ell_m$ is placed at the back focal plane of L6 to implement topological charge matching. Finally, the output beam is focused by L7 and imaged by a CMOs camera (CCD) on the focal plane. The SPP, L7, and CCD collectively form the OAM detection system. As previously mentioned, the effectiveness of our scheme will be verified based on the results of topological charge matching captured by the CCD. Besides, during the detection of the intensity distribution, the SPP and L7 are removed, and the CCD is relocated on the back focal plane of L6 to directly capture the image of the output beam.

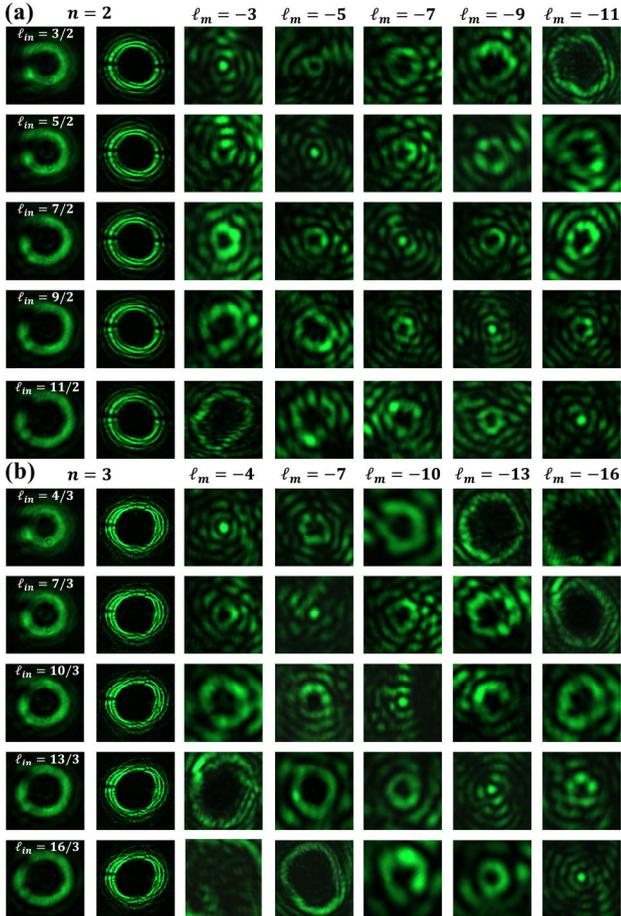

FIG. 4. Experimental results of effective sorting of two sets of FOAM modes with different parameters (a) $t = 1/2$ and $n = 2$, (b) $t = 1/3$ and $n = 3$. The first two columns are the intensity distribution before and after the transformation, respectively.

Figure 4 summarizes the experimental results of effective sorting of fractional vortex modes. Our sorting method is effective for any set of FOAM modes with the same $t$. To verify this, we conduct sorting experiments on two sets of fractional vortex modes. Figure 4(a) shows the results for FOAM modes with $t = 1/2$ and $n = 2$, including $\ell_{in} = 3/2, 5/2, 7/2, 9/2, 11/2$. Figure 4(b) shows the results for FOAM modes with $t = 1/3$ and $n = 3$, including $\ell_{in} = 4/3, 7/3, 10/3, 13/3, 16/3$. The corresponding transformation phase and correction phase are illustrated in Figs. 3(a)-3(b). The first two columns in each group are the intensity distribution before and after the transformation. As expected, the spiral vortex distribution is observed in all groups. The results of sorting by topological charge matching also match well with the numerical simulations. The $\ell_m$ of the vortex maps used are equal to $-n\ell_{in}$; i.e., from $-3$ to $-11$ in Fig. 4(a) and from $-4$ to $-16$ in Fig. 4(b). It can be clearly seen that for each FOAM input mode, only when the $\ell_m$ and the target topological charges $\ell_{out}$ match each other ($\ell_{all} = 0$), the intensity distribution obtained on CCD is a single Gaussian-like spot in the center of the field. Under the condition of mismatching, the intensity pattern is a halo larger than the Gaussian-like spot. The agreement of the experimental results with the simulations is quite satisfactory in terms of the shape and the position of the results of topological charge matching, confirming the successful conversion from FOAM modes to IOAM modes. In each column, there is almost no overlap between the Gaussian-like spot and the halos, indicating that our scheme allows for effective sorting with low crosstalk of such a set of FOAM modes by using spatial filter to extract the central Gaussian-like spot. It is also noted that the patterns are surrounded by some stray light in the background, which is attributed to the imperfect phase matching in the process caused by small errors in experimental setup, such as transverse displacement and the distance between SLMs.

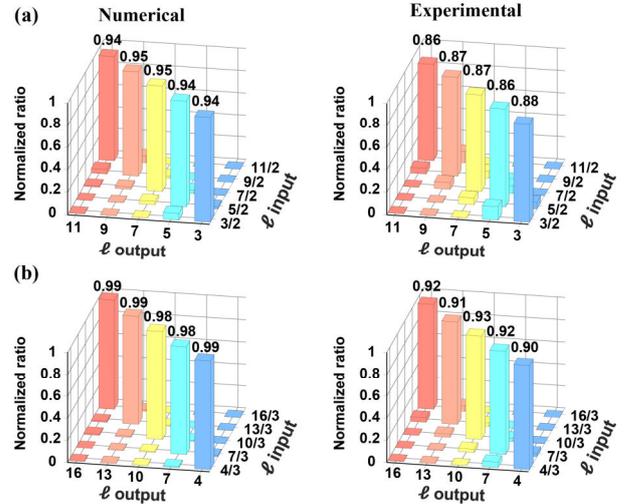

FIG. 5. The numerical and experimental OAM mode purity of the sorted FOAM modes with (a) $t = 1/2$ and $n = 2$, (b) $t = 1/3$ and $n = 3$.

Further results demonstrating the ability of our scheme for effective sorting of FOAM modes is the OAM mode purity analysis of the extracted beam. Following the principle of topological charge matching, in the case of using the same matched vortex phase map of $\ell_m$, we can determine the mode purity by measuring the intensity in the region of the central Gaussian-like spot for all the input states. The purity of different input states in the extracted beam is the ratio of each measured intensity to the total measured intensity. Figure 5 shows the numerical and experimental results for both sets of FOAM modes. Since there is little overlap between the patterns of two neighboring FOAM states, the results indicate high purity of the target FOAM mode in the extracted beam, which verifies the success of effective sorting with low crosstalk of fractional vortex modes.

In the present form, our method has demonstrated its powerful ability to effectively sort arbitrary FOAM modes. However, the efficiency in our method is limited by the topological charge matching method, which can only extract one FOAM mode at a time. Such limitation can be improved by combining other efficient IOAM sorting methods [25,26], indicating the potential of our method to further realize the efficient sorting of FOAM modes. It is worth mentioning that the communication system based on FOAM modes has significant advantages in terms of confidentiality, as they are more difficult to identify and sort compared to IOAM modes. In this regard, our method can serve as the encryption key to implement highly confidential FOAM-based communication systems. In addition, since the coordinate transformation method has no wavelength restriction, our method can also be generalized to other wave bands.

In conclusion, we have theoretically and experimentally reported a novel scheme based on coordinate transformation to implement the effective sorting of fractional vortex modes, for the first time to our best knowledge. Starting from the theoretical predictions, we propose a coordinate mapping method to convert FOAM non-eigenmodes to IOAM eigenmodes, after which the effective sorting of different FOAM modes can be implemented by topological charge matching. The theoretical predictions have been verified through numerical simulations and experiments. Furthermore, the OAM mode purity analysis confirms that the sorting of FOAM modes has low crosstalk. With the ability to perform effective sorting of fractional vortex modes, our work paves the way for the FOAM-based communication. More attractively, our scheme can further provide encryption for the communication process, which will open up new applications in both the classical and the quantum regimes.

We wish to acknowledge the support of National Natural Science Foundation of China (NSFC) (12192252 and 12374314); National Key Research and Development Program of China (2013YFA1407200).

---

* These authors contributed equally to this work.
† liuhaigang@sjtu.edu.cn
‡ xfchen@sjtu.edu.cn